\documentclass[twocolumn,showpacs,preprintnumbers,amsmath,amssymb,floatfix]{revtex4-1}
\usepackage{graphicx,epsf,epsfig}% Include figure files
\usepackage{subfigure}
\usepackage{float}
\usepackage{dcolumn}% Align table columns on decimal point
\usepackage{bm}% bold math
\usepackage{ulem}
\usepackage{setspace}

\begin{document}

%\preprint{}

\title{High efficiency switching using graphene based electron `optics'}
\author{Redwan N. Sajjad and Avik W. Ghosh}
\affiliation{Department of Electrical and Computer Engineering \\
University of Virginia, Charlottesville, VA 22904 \\
}%
\date{\today}% It is always \today, today,
             % but any date may be explicitly specified

%\pacs{72.80.Vp, 73.63.-b, 72.10.-d}% PACS, the Physics and Astronomy

\begin{abstract}
The absence of a band-gap in graphene limits the gate modulation of its electron conductivity, both in regular graphene as well as in PN junctions, where electrostatic barriers prove transparent to Klein tunneling. We demonstrate a novel way to directly open a gate-tunable transmission gap across graphene PN junctions (GPNJ) by introducing an additional barrier in the middle that replaces Klein tunneling with regular tunneling, allowing us to electrostatically modulate the current by several orders of magnitude. The gap arises by angularly sorting electrons by their longitudinal energy and filtering out the hottest, normally incident electrons with the tunnel barrier, and the rest through total internal reflection. Using analytical and atomistic numerical studies of quantum transport, we show that the complete filtering of all incident electrons causes the GPNJ to act as a novel metamaterial with a unique gate-tunable transmission-gap that generates a sharp non-thermal switching of electrons. In fact, the transmission gap gradually diminishes to zero as we electrostatically reduce the voltage gradient across the junction towards the homogeneous doping limit. The resulting gate tunable metal-insulator transition enables  the electrons to overcome the classic room temperature switching limit of $k_BTln10/q \approx 60$ mV/decade for subthreshold conduction.
\typeout{polish abstract}
\end{abstract}
\maketitle

\begin{figure*}[!htp]
\centering
%\hskip 0.1cm\includegraphics[width=3.2in]{Figures/journal/figure1a.png}\quad
%\hskip-2.5cm\subfigure{\includegraphics[width=3.2in,height=1.3in]{Figures/journal/figure1b.png}}
\includegraphics[width=3.2in]{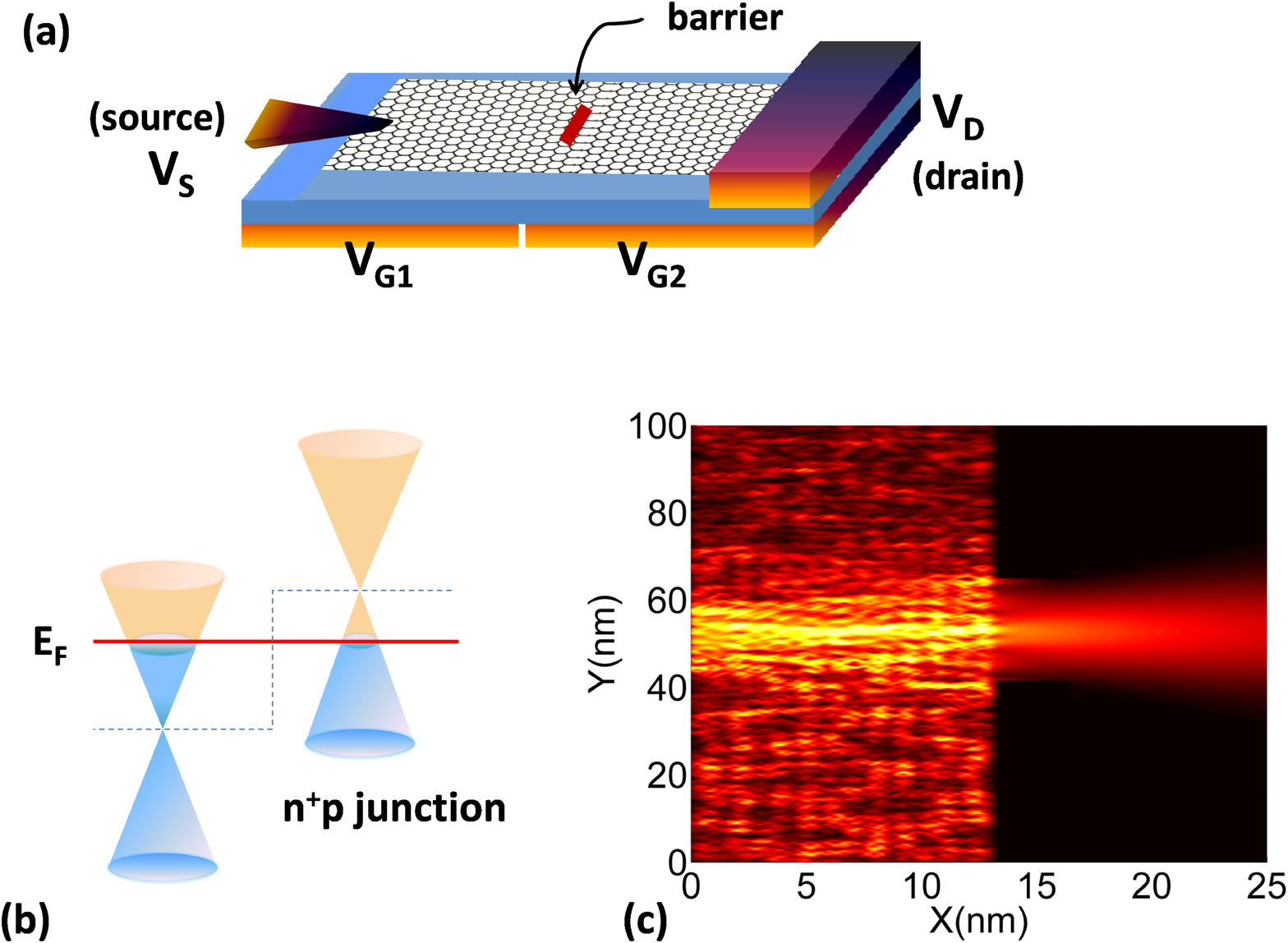}\quad
\subfigure{\includegraphics[width=3.2in]{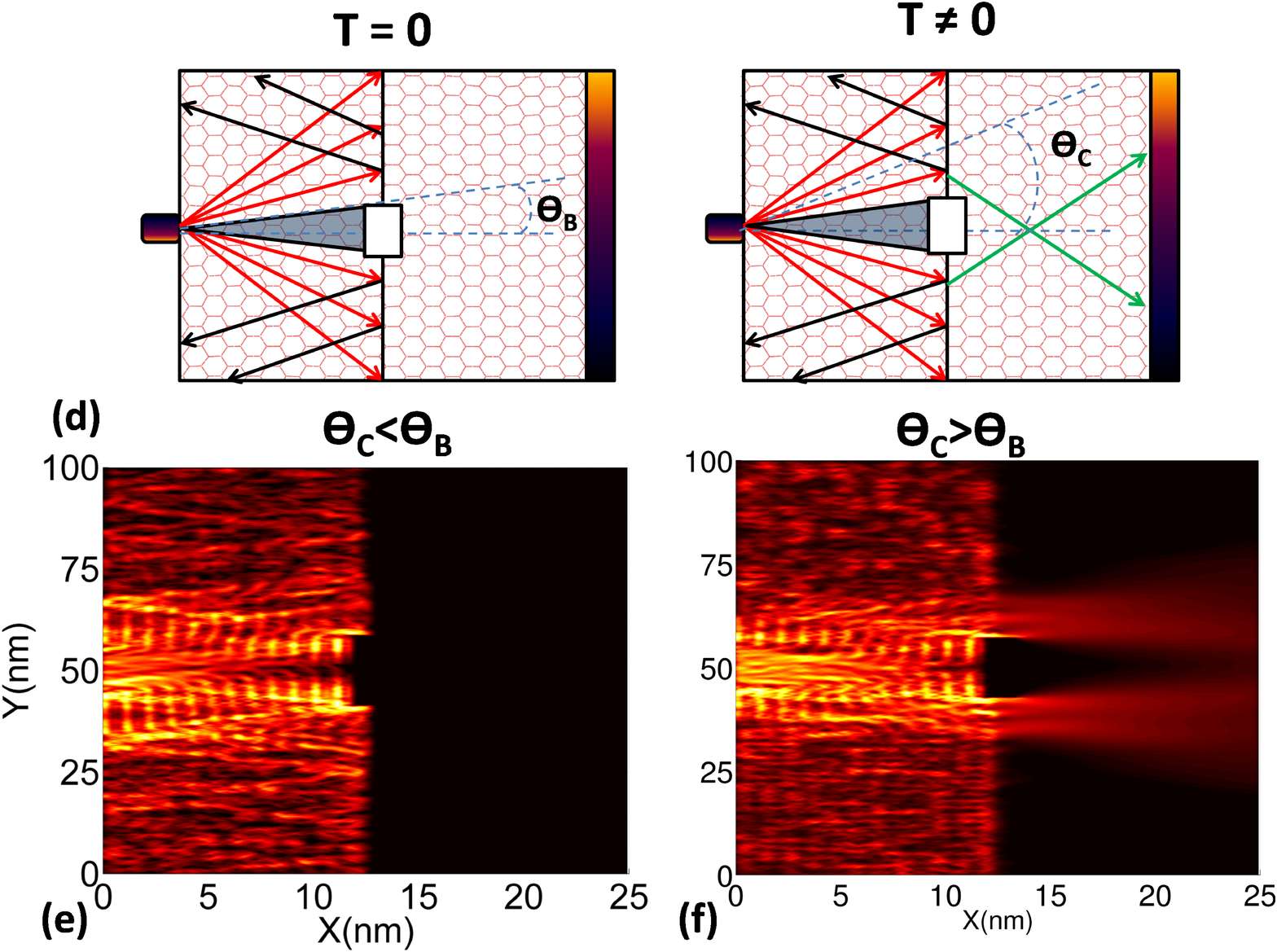}}
%\hskip -0.7cm\subfigure{\includegraphics[width=3.5in height=0.3in]{Figures/journal/fig1c_dev.png}}
%\caption{(Top) A point source angularly spreads electrons while an extended drain collects those that
%refract around a barrier. Split gates control the two graphene segments separately.
%Schematic (middle) as well as atomistic NEGF plots (right) illustrate how the transmission turns
%off at a gate voltage $V_{G2}$ where the critical angle $\theta_C < \theta_B$, where $\theta_B$ is the
%occlusion angle created by the barrier. The ON current 
%corresponds to $\theta_C >\theta_B$ where electrons can refract around the barrier. The result
%is a voltage-dependent transmission gap (Figs. 3,4) creating a sharp non-thermal switching at
%less than the textbook limit of 60mV/decade (Fig.4)}\label{device}
\caption{(a) A point source angularly spreads electrons while an extended drain collects those that
refract around a tunnel barrier. Split gates control the two graphene segments separately.
For the energy band diagram (b) with $|$V$_{G1}|$$>|$V$_{G2}$$|$ and very small drain bias, (c) an atomistic current density plot calculated with the Non-Equilibrium Green's function (NEGF) formalism  illustrates that current flows only within the critical angle,  $\theta_C = sin^{-1}(V_{G2}/V_{G1})$ due to direct band-to-band (Klein) tunneling. (d) Schematic Snell's Law predictions as well as (e) fully atomistic quantum calculations illustrate how a barrier eliminates transmission within the critical angle when $\theta_C < \theta_B$, where $\theta_B$ is the occlusion half angle created by the barrier. The non zero transmission occurs only for $\theta_C >\theta_B$ where electrons can refract around the barrier (f). The elimination of transmission by coupling angular filtering with energy filtering  ultimately generates the high switching efficiency of GPNJs}\label{device}
\end{figure*}

Interest in graphene electronics is largely driven by its impressive material and electronic properties \cite{novoselov_04,katsnelson_06,zhou_06}. The two graphene bands derive from bonding-antibonding combinations of neighboring carbon $p_z$ dimer (`pseudo-spin') basis sets belonging to the same two-dimensional crystallographic point group.  The corresponding gapless low-energy excitations generate an ultralow electron effective mass, while the orthogonality of the pseudo-spins suppresses 1-D back-scattering \cite {ando_98}, resulting in an incredibly high electron mobility \cite{liao_10,morozov_08, bolotin_08}. Unfortunately, the gaplessness also makes its switching properties quite modest \cite{schwierz_10}, the subthreshold current changing linearly rather than exponentially with voltage. Conversely, opening a bandgap {\it{structurally}} by chemical modifications or quantum confinement \cite{guo_10,son_06,zhang_09} fundamentally reduces the mobility due to an asymptotic constraint on its short wavelength behavior \cite{tseng_10}, rendering high efficiency switching in graphene a considerable challenge.

The phase correlation between conduction and valence band states allows uniquely novel forms of electron flow in graphene. At a graphene p-n junction, the scattering of the individual pseudospins leads
to electron trajectories that are reminiscent of E-M wave scattering at dielectric interfaces. Since the radius of the 2-D Fermi surface varies with local potential across an interface, electrons with large angles of incidence at the higher electrostatically doped side are unable to conserve their transverse quasi-momentum components across the junction and thus undergo total internal reflection. Those within a critical angle `refract' into the opposite side, diverging if the doping has the same sign (i.e. n$^+$n or p$^+$p junctions), or converging if opposite (pn junctions). A varying electrostatic doping thus turns graphitic electrons into quantum mechanical analogues of negative index metamaterials \cite{ref1}, the linear E-k making the electron trajectories non-dispersive over a finite energy window imposed by temperature and drain bias. Such electron `optics' allows us to modulate the conductivity of a graphene sheet with split gates \cite{low_09,xing_10,williams_07,huard_07} or build electronic analogues of fiber optic cables \cite{low_11}. However, normally incident electrons well below the critical angle can directly tunnel between bands (`Klein tunneling') even for large voltage gradients, once again reducing the ON-OFF ratio.

\begin{figure}
\includegraphics[width=3.3in]{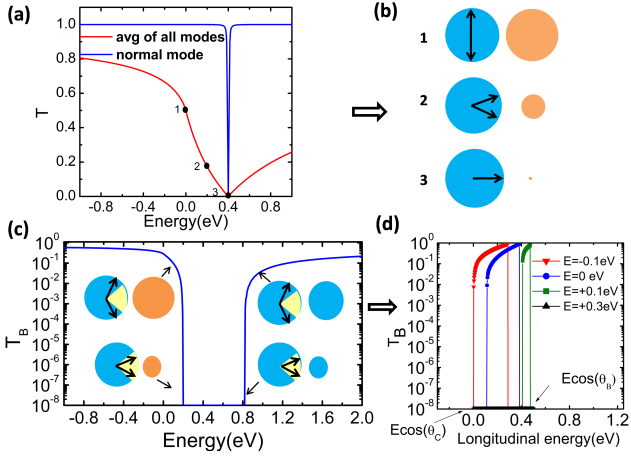}
%%\subfigure{\includegraphics[width=1.7in]{Figures/journal/graph13.png}}
\caption{(a) GPNJ Transmission vs. Fermi energy from Eq.~\ref{tr_theta} for $V_{G1}$=+0.4V and $V_{G2}$=-0.4V. At all energies, except when aligned with the Dirac point on the p side, the zero angle mode transmits with unit probability, making the average transmission nonzero. (b) The circles represent constant Fermi energy slices of the Dirac cones across the junction, corresponding to points on the $T(E)$ curve. Arrows indicate the maximum $k_{y}$ values that transmit across the junction, and thus define the critical angle. (c) Modification of the transmission spectrum from part (a) after putting the barrier; transmission within the yellow region is forbidden due to barrier. (d) Replotting T vs longitudinal energy $E_L$ at specific total energies $E$, indicating an effective band-pass behavior for states outside the $T(E)$ transmission gap.}
 \label{graph1}
\end{figure}

\begin{figure}[h!]
\vskip -0.1in
\hskip -0.2in\includegraphics[width=3.8in]{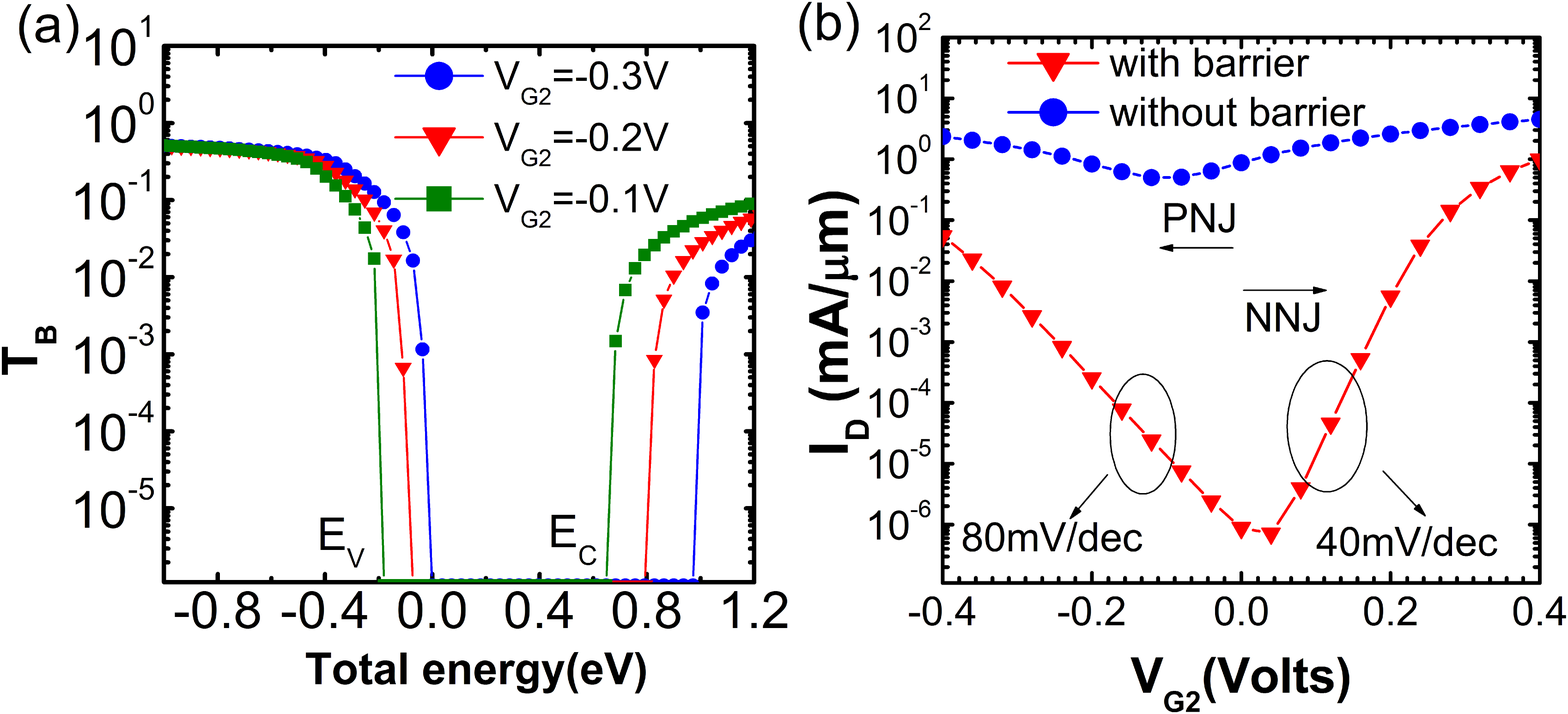}
%\centering
\vskip -0.3in\caption{(a) Transmission vs total energy for different gate voltages. The transmission shifts as we 
vary the voltage $V_{G2}$, but is accompanied with a change in transmission gap; (b) I-V$_{G2}$ of the device, 
showing a sharp increase in current modulation with barrier, and a subthreshold swing that is larger than the textbook limit
at the PN end and smaller at the N$^+$N end. Here, V$_{G1}$ =+1volt, $\theta_B$ = 20$^0$ and a drain voltage of 0.2V.}
\label{graph4}
\end{figure}

In this letter, we explore a graphene junction where electrons injected by a point source are spectrally separated by a local gate, and those refracted across the junction are collected with an extended drain (Fig. 1). When electrons are injected from a higher electrostatically `doped'  to a lower doped side, the ones with high incident angle, low longitudinal energy are eliminated by total internal reflection. An additional tunnel barrier removes the normally incident, hotter electrons as well, resulting in a \textit{voltage-dependent transmission gap} that allows significant gate modulation of the electron current (Figs.~\ref{graph1}-\ref{graph4}). Furthermore, as we vary the gate voltage on the drain side, moving progressively towards a homogenous doping across the junction, we see an effective {\it{upconversion}} of the local voltage, leading to a subthreshold swing lower than the textbook thermal limit of $k_BT\ln{10}/q$. The reduction in  swing, that ultimately limits low energy switching, arises from the voltage tunability of the transmission gap, and in fact collapses the bandgap in the homogeneous doping limit, requiring the `valence band-edge' to slow down while the `conduction band-edge' to catch up. We thus have a unique non-thermal switching mechanism compared to other low subthreshold swing devices proposed \cite{sayeef_08,kailash_02,appenzeller_04}, arising specifically from a gate-tunable metal-insulator transition that bypasses the traditional limitations of tunnel based switches such as low ON current.

{\it{Normal transmission in a GPNJ.}} The wavefunctions for graphitic electrons follow from the Dirac Hamiltonian, $H(\vec{k}) = \hbar v_F\vec{\sigma}.\vec{k}$,
where $\vec{\sigma} = (\sigma_x, \sigma_y)$ are the Pauli matrices and $v_F \approx 10^6 ms^{-1}$ is the Fermi velocity for electrons in graphene. The wave-functions consist of a plane wave part with Fermi wavevector magnitudes $k_{F1,F2}$ oriented along incident ($\theta_{1}$), reflected ($\theta_{1}$) and transmitted angles ($\theta_{2}$) on the two sides, while their Bloch-like atomic parts act as two-component spinors whose phases are determined by the angles and overall signs determined by the band-indices. Matching the plane wave phases across the boundary, we get the corresponding Snell's Law $k_{F1}sin\theta_1=-k_{F2}sin\theta_2$ with a refractive index ratio $n=-{k_{F2}}/{k_{F1}}$. Matching the spinor (Bloch) components thereafter gives us the transmission probability
\begin{eqnarray}\label{tr_theta}
T(\theta_1,E, V_G)=\Theta(\theta_C-\theta_1)\frac{cos\theta_1 cos\theta_2}{cos^2\biggl(\displaystyle\frac{\theta_1+\theta_2}{2}\biggr)}
\end{eqnarray}
where $\Theta$ is the unit step function, $\theta_2$ is related to $\theta_1$ by Snell's law, $V_G$=$\{V_{G1}, V_{G2}\}$ is a particular set of gate voltages and $\theta_C = sin^{-1}(n)$ is the critical angle. The equation gives zero transmission (total internal reflection) for $\theta_1\geq\theta_C$, unit transmission for a homogenous sheet ($\theta_1=\theta_2$), and focusing with $T(\theta_1)=cos^2{\theta_1}$ for symmetric GPNJs ($\theta_1=-\theta_2$).
The transmission can also be plotted vs. longitudinal energy $E_L$ for a given total energy $E = \sqrt{E_L^2 + E_T^2}$. The  critical angle $\theta_C$ eliminates the obliquely incident electrons with high transverse and low longitudinal energy, so that $T(E_L)$ looks like a {\it{high-pass filter}}, transmitting only states with $E_L > E\cos{\theta_C}$.

The energy dependence of T enters through the $\theta$ terms related by Snell's law,
while the gate voltage dependence enters through the positions of the charge neutrality points on both sides of the junction. The transmission here is that of a single mode $\{k_x,k_y\}$ with incident angle $\theta_1 = tan^{-1}({k_y}/{k_x})$ and energy $E = \hbar v_F\sqrt(k_x^2+k_y^2) = \sqrt{E_L^2+E_T^2}$. The average transmission of all modes at a given total energy $E$ is found by integrating T from Eq. \ref{tr_theta} over all incident angles and then normalizing (Fig. \ref{graph1}a). As E approaches the neutrality point on the refracted side, more and more modes suffer total internal reflection, until at $-qV_{G2}$ (0.4 eV in this case), the critical angle vanishes and all  modes are reflected. Since this happens at a single voltage where all incident modes align with the Dirac point on the refracted side, the ON-OFF ratio is quite poor and in fact comparable to a regular graphene transistor. In effect,  electrons are injected over a window of energy  set by the drain bias, creating a finite cone over which transmission will take place (Fig.~\ref{device}c) and degrading the ON-OFF. Since electrons with higher longitudinal energy see a lower barrier, {\it{eliminating transmission from the high longitudinal energy electrons is critical to higher efficiency}} of the Klein tunnel switch.

{\it{Creating a transmission gap.}} Fig. \ref{device}a shows the schematic diagram of the device we propose. The source injects electrons at all angles. If we pattern a barrier at the center of the graphene sheet, the normally incident electrons will be eliminated, so that the transmission probability with barrier can be written as
\begin{equation}\label{tr_b}
T_B(\theta_1,E, V_G) = \Theta(\theta_1-\theta_B) T(\theta_1, E, V_G)_{}
\end{equation}
where $\theta_B = tan^{-1}{D}/{2L}$ is the occlusion half angle subtended at the source by the barrier,  D is length of the rectangular barrier and L is its perpendicular distance from the source. Atomistic NEGF calculations in Fig. \ref{device}e confirm how modes near normal incidence are reflected by the barrier, while those incident at a large angle are eliminated by total internal reflection. The transmission now acts as a {\it{band-pass filter}} along the longitudinal energy axis $E_L$ (Fig. \ref{graph1}d), allowing only electrons with angles $\theta_B < \theta_1 < \theta_C$, in other words, within a longitudinal energy window $E\cos{\theta_C} < E_L < E\cos{\theta_B}$ to transmit.
Since no electrons can transmit when $\theta_C < \theta_B$, there is a range of energies for which we get a transmission gap. The gate voltage dependance of $\theta_C$ leads thereby to a {\it{voltage-tunable transmission gap}}.
%\begin{figure}%[!t]
%\centering
%\includegraphics[width=3.6in]{Figures/journal/transm.png}\quad
%%%\subfigure{\includegraphics[width=2in]{Figures/focusing_cut.jpg}}
%\caption{Schematic diagram of the device,  (a) total internal reflection (b) focusing.} \label{graph3}
%\end{figure}

{\it{Gating the transmission gap: a continuous metal-insulator transition.}}
We have shown that filtering of electrons with high longitudinal energy leads to a transmission gap.
One can extract an effective `valence band-edge' $E_V$ when this happens, given by the condition 
$sin\theta_B = sin\theta_c =-{(E_V+qV_{G2})}/{(E_V+qV_{G1})}$. One can get a similar expression for the `conduction band-edge' $E_C$, and thus the effective transmission gap $E_G$
\begin{eqnarray}\label{en_tr1}
E_{V,C}& =&-qV_{G2} \mp q\frac{(V_{G1}-V_{G2})\sin\theta_B}{1\pm sin\theta_B}\nonumber\\
E_G &=& E_C - E_V = q(V_{G1}-V_{G2})\frac{2sin\theta_B}{cos^2\theta_B}
\end{eqnarray}
Note that there are two distinct contributions to the resulting transmission plot (Fig.~\ref{graph4}a)-- 
(i) an overall shift given by $-qV_{G2}$, and (ii) a voltage modulation of the band-gap
$E_G$. The first term will lead to the electrostatic gating effect seen for regular 
band-gapped semiconductors (for good gate control geometries, this amounts
to $k_BT\ln{10}/q \approx 60$ mV/decade), while the latter term will give a deviation from this textbook
result, effectively captured through a gate tunable, continuous metal-insulator transition
associated with voltage dependence of the transmission gap $E_G$. As expected, 
the band-gap vanishes in the absence of either a tunnel barrier ($\theta_B = 0$), or a
voltage gradient across the junction ($V_{G2} = V_{G1}$, $\theta_C = 90^0$).

%%\begin{figure*}%[h!]
%%\includegraphics[width=6in]{Figures/journal/graph4.png}
%%\centering
%%\caption{ }
%%\label{graph4}
%%\end{figure*}

Fig.~\ref{graph4} shows the current extracted by integrating the analytical transmission formula (Eq.~\ref{tr_b}) using the Landauer
formalism  for various gate voltages $V_{G2}$ while $V_{G1}$ is kept fixed. While the ON current reduces a little, 
the tunnel barrier primarily reduces the OFF current (Fig. 4), dropping it by nearly five orders of 
magnitude. Overall, an ON-OFF ratio of $\sim 10^4$ is achievable with just 0.25 V change in gate voltage 
on the NN regime with $\theta_B$ of $20^0$. 

The unusual transconductance and subthreshold swing in Fig.~\ref {graph4} can attributed to the fact that transmission gap changes everytime the gate voltage $V_{G2}$ is changed. Eq.\ref{en_tr1} gives us the edges of the transmission gap, so 
${dE_{C,V}}/{dV_{G2}} = {\alpha_g q}/{[1\mp sin\theta_B]}$, where we have now introduced the capacitive gate transfer factor $\alpha_g$. In other words, the subthreshold swing,ultimately dictating the efficiency of low voltage switching, will be
\begin{equation}
S = ({d log_{10}I}/{dV_{G2}})^{-1} = \frac{k_BT\ln{10}}{\alpha_gq}[1\mp sin\theta_B]
\end{equation} which is less than 60mV/dec (`-' sign above) at the $NN$ end, and more than 60mV/dec (`+' sign) at the $NP$ end.
At heart of the unusual behavior is the coupling of angular and energy filtering in the barrier driven Klein tunnel-switch, which differs from most materials in that {\it{its effective band-gap can be collapsed with a gate voltage}}. 

The tunability of transmission gap is reproduced with a fully atomistic NEGF tight-binding calculation of current flow across a GPNJ sheet involving $\sim$ 100,000 atoms (Figs. 1, 4). The transmission gap is slightly lower than the analytical predictions owing to the presence of leakage currents created by edge reflection of the electron waves. Such effects are expected to be substantially reduced if we edge passivate large sheets of graphene. Indeed, selectively imposing absorbing self-energy matrices minimizes such standing waves, causing the computed
transmission gap to approach our analytical predictions. 

The physics of regular tunneling coupled with Klein tunneling opens up the opportunity for high performance low power switching based on graphene. Our results are predicated upon Snell's Law, in other words, the preservation of the transverse quasi-momentum across the junction. Various non-idealities can compromise these effects, such as width-independent leakage currents arising from edge reflections and diffraction effects at the barrier edges. These non-idealities are already accounted for by our numerical calculations and dominate primarily for long wavelength electrons that are, in fact,  eliminated by the tunnel barrier. 
For a 1$\mu m \times 1 \mu m$ sheet with an occlusion half angle $\theta_B = 20^0$, a $\sim 180~nm \times 1 nm$ rectangular void generates a tunneling probability that is considerably less than 10$^{-8}$. Experimental success depends on the mobilities of graphene samples, and the quality of the tunnel barrier created either using electron beam lithography or by directed assembly of insulating molecular species on graphene, as demonstrated by Hersam et al \cite{wang_09}. Further numerical simulations are needed to study the robustness of these effects in the face of charge puddles that could misalign the Dirac points within each graphene segment, finite size effects at the point contact, line-edge roughness at the electrodes, finite junction widths imposed by the Debye lengths, and pseudospin mixing by edge states with various degrees of chemical passivation.

{\it{Acknowledgments.}} We thank Ji Ung Lee, Tony Low, Mark Lundstrom, Frank Tseng, Golam Rabbani, Sayeef Salahuddin and Mircea Stan for useful discussions. This work was supported by the NRI grant within the INDEX center. 

\bibliographystyle{apsrev4-1}
%\bibliography{sajjad_library}
%
\end{document}